\documentclass[prd,twocolumn,showpacs,superscriptaddress,groupedaddress,nofootinbib]{revtex4-1}

\usepackage{graphicx}
\usepackage{amsmath,amssymb} 
\usepackage{color} 
\usepackage{xspace}

\usepackage{tikz}
\usetikzlibrary{arrows,automata}

\newcommand{\PlotPath}{./}

\newcommand{\Hypo}{\ensuremath{H}\xspace}

\begin{document}


\title{An information-theoretic approach to the gravitational-wave burst detection problem}
\author{Ryan Lynch}\email{ryan.lynch@ligo.org}
\author{Salvatore Vitale}
\author{Reed Essick}
\author{Erik Katsavounidis}
\affiliation{Massachusetts Institute of Technology, 185 Albany St, 02139 Cambridge USA}
\author{Florent Robinet}
\affiliation{LAL, Univ. Paris-Sud, CNRS/IN2P3, Université Paris-Saclay, Orsay, France}


\begin{abstract}

The observational era of gravitational-wave astronomy began in the Fall of 2015 with the detection of GW150914.  
One potential type of detectable gravitational wave is short-duration gravitational-wave bursts, whose waveforms can be difficult to predict.  
We present the framework for a new detection algorithm for such burst events -- \textit{oLIB} -- that can be used in low-latency to identify gravitational-wave transients independently of other search algorithms.  
This algorithm consists of 1) an excess-power event generator based on the Q-transform -- \textit{Omicron} --, 2) coincidence of these events across a detector network, and 3) an analysis of the coincident events using a Markov chain Monte Carlo Bayesian evidence calculator -- \textit{LALInferenceBurst}.  
These steps compress the full data streams into a set of Bayes factors for each event; through this process, we use elements from information theory to minimize the amount of information regarding the signal-versus-noise hypothesis that is lost.  
We optimally extract this information using a likelihood-ratio test to estimate a detection significance for each event.   
Using representative archival LIGO data, we show that the algorithm can detect gravitational-wave burst events of astrophysical strength in realistic instrumental noise across different burst waveform morphologies.  
We also demonstrate that the combination of Bayes factors by means of a likelihood-ratio test can improve the detection efficiency of a gravitational-wave burst search.  
Finally, we show that oLIB's performance is robust against the choice of gravitational-wave populations used to model the likelihood-ratio test likelihoods.

\end{abstract}
\maketitle

\section{Introduction}\label{Sec.Intro}

With the first detections of gravitational waves (GWs) \cite{GW150914,BoxingDay}, gravitational-wave astronomy has blossomed into an observational field. 
The two Advanced LIGO detectors \cite{aLIGO} -- one in Livingston, LA and the other in Hanford, WA -- conducted their first observing run between September 2015 and January 2016, and Advanced Virgo \cite{aVirgo} is expected to join them in 2017.
These advanced detectors are expected to reach their design sensitivities within the next 2-3 years \cite{ObservingScenario}.
Two additional instruments, LIGO India \cite{INDIGO} and the Japanese KAGRA \cite{Kagra} should join the global network before the end of the decade, further increasing the sensitivity to GWs.

There are many potential astrophysical sources that could be observed by these instruments.
Some, such as the inspiral and merger of compact stellar remnants, known as compact binary coalescence (CBC), have well-modeled and well-understood theoretical waveform predictions (see e.g. Refs~\cite{Hannam:2013oca,2016PhRvD..93d4007K,2016PhRvD..93d4006H,2016arXiv160601210T} for a description of the waveforms used to analyze the events detected by LIGO in the first observing run).
With these models in hand, Weiner matched-filtering techniques provide optimal detection schema. 
Extensive effort goes into continuously improving these models (compare e.g. the subsequential versions of the SEOBNR~\cite{2014PhRvD..89h4006P} or IMRPhenom~\cite{Hannam:2013oca} waveforms) and compare them with numerical relativity simulations~\cite{2016arXiv160601262T}.
These efforts have already contributed to two high-confidence detections of binary-black-hole mergers \cite{GW150914,BoxingDay}.
However, there are other types of GW sources with poorly modeled or unknown waveforms, such as core-collapse supernovae \cite{Ott:2006,Ott:2009,Burrows:2006,Burrows:2007,Dimmelmeier:2008}, magnetar-induced neutron star glitches \cite{SGRR,Magnetar}, and cosmic string cusps \cite{CosmicStrings}.
This paper focuses on short duration ($\leq 1$ second) unmodeled transients with frequencies between ten Hz and a few kHz, commonly know as GW bursts.
Therefore, throughout this paper we make the assumption that the targeted signals are inherently unknown in origin and morphology, although searches for unmodeled bursts are indeed sensitive to the better understood sources mentioned above \cite{IMBH}.
This sensitivity was explicitly validated with the detection of GW150914 \cite{BCP}.

Discovering unmodeled sources of GWs is an exciting prospect for the advanced detectors. 
In particular, localizing generic sources in the sky \cite{BF2Y} could provide information about their origin, and accurate reconstruction of the waveform could determine their emission mechanism, which is especially promising for supernovae \cite{SMEE}. 
However, before this information is available or robust enough, we must ensure that we have confidently detected a GW signal.
In this way, we can separate in-depth parameter estimation from detection.
This paper focuses primarily on the detection problem and presents a new algorithm -- \textit{oLIB} -- that generates significance estimates for GW burst candidates via nearly-lossless compression of the information contained within the raw data.

Many different burst detection statistics and end-to-end search algorithms have been used historically \cite{S6Burst,Klimenko:2005,Klimenko:2008,Klimenko:2011,Sutton:2010,BWcomplexity}.
In particular, another algorithm \cite{BayesWave, BayesLine} has recently claimed the ability to make high-confidence detections \cite{BWcomplexity} using the Bayesian evidence computed by a stochastic sampler as a follow-up to other search algorithms. 
This approach is similar in scheme to the end-to-end oLIB algorithm.
It is of great interest to have overlap between multiple search algorithms so that cross-validation can be carried out for detection candidates.
Although most detection schemes are motivated by similar noise models for the detectors, which typically assume stationarity and Gaussianity, there is uncertainty regarding how optimal their exact search statistics are for unmodeled bursts in real non-Gaussian detector noise.  
This paper presents a method for algorithmically generating optimal search statistics for proposed signal and noise hypotheses through an application of information-theoretic concepts.  
This method then compresses these search statistics into a single, scalar search statistic.
This compression is done in such a manner that it minimizes the information lost concerning the signal-versus-noise hypothesis.

oLIB is an attempt to implement this optimal scheme.  The implementation is carried out by first flagging, in each detector, subsets of data that have excess power, which we refer to as ``events''.
This step is carried out with a time-frequency decomposition based on the Q-transform \cite{Gabor, Brown, Qtransform} that we will refer to as \textit{Omicron} \cite{Omicron}.  
This first step is followed by a time coincidence of such excess power among the network of detectors.
The resulting set of coincidences are handed to a follow-up, performed with LALInference Burst (LIB) \cite{LALInference_nest,LALInference,BF2Y}, that analyzes all data streams simultaneously and compresses them into a set of Bayes factors.
Applying a likelihood-ratio test (LRT) to these Bayes factors produces a single search statistic, which is then mapped into an estimate of the GW detection significance.
At each step in the algorithm, we take care to analyze possible losses of information, which include modeling uncertainty and waveform mismatch, among other sources.

Although the signals that oLIB targets are inherently unknown, the algorithm must make some limited assumptions regarding their morphology.  
oLIB is more sensitive to signals that better match these assumptions, but it can still detect generic signals at astrophysically relevant signal amplitudes that differ significantly from its internal models.
Furthermore, these robust detection statements can be reached in real time, allowing oLIB to initiate and inform the rapid electromagnetic follow-up of GW candidates.  
GW150914 proved this, with oLIB being one of two independent search algorithms to detect the event in low-latency \cite{BCP}.

We describe oLIB's algorithmic structure in more detail in \S \ref{Sec.Description}.
Using archival (public) LIGO data, we present a proof-of-concept analysis in \S \ref{Sec.Application}, which is meant to validate the design choices of the algorithm.
Finally, we conclude with a summary in \S \ref{Sec.Conclusions} and provide some technical details in the Appendix.

\section{Algorithm Description}\label{Sec.Description}

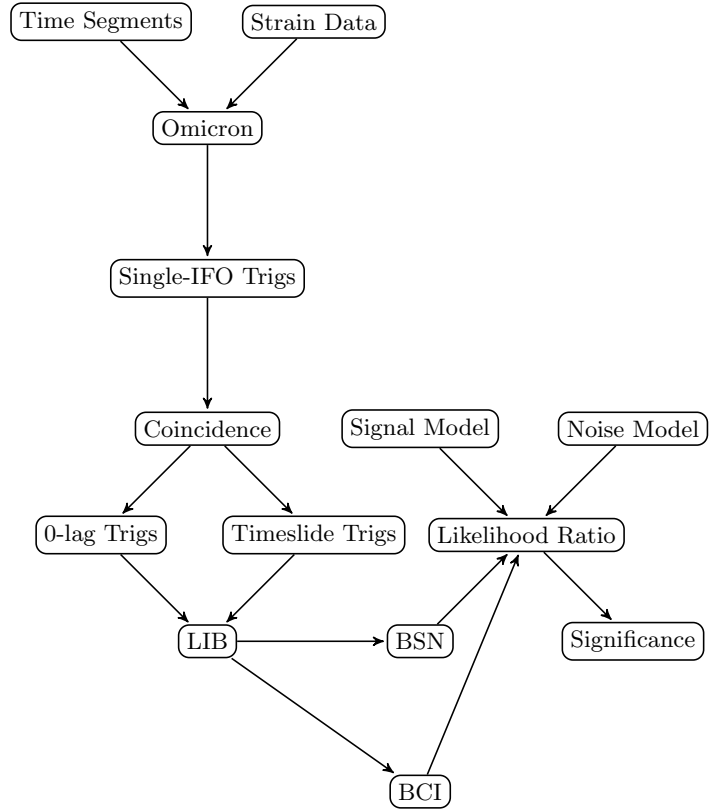
\begin{figure}[h!]
    \begin{tikzpicture}[->,>=stealth', shorten >=1pt, auto, node distance=2.00cm, semithick]
        \tikzset{
           data/.style={
                        rectangle,
                        rounded corners,
                        draw=black,
                        fill=none,
                        text=black
                       },
           actn/.style={
                        rectangle,
                        rounded corners,
                        draw=black,
                        fill=none,
                        text=black
                       },
           stat/.style={
                        rectangle,
                        rounded corners,
                        draw=black,
                        fill=none,
                        text=black
                       },
           rslt/.style={
                        rectangle,
                        rounded corners,
                        draw=black,
                        fill=none,
                        text=black
                       }
                 };

        \node[data]         (segs)                             {Time Segments};
        \node[actn]         (omicron) [below right of=segs]    {Omicron};
        \node[data]         (frames)  [above right of=omicron] {Strain Data};
        \node[data]         (sngls)   [below of=omicron]       {Single-IFO Trigs};
        \node[actn]         (coinc)   [below of=sngls]         {Coincidence};
        \node[data]         (0-lag)   [below left of=coinc]    {0-lag Trigs};
        \node[actn]         (LIB)     [below right of=0-lag]   {LIB};
        \node[data]         (slides)  [below right of=coinc]   {Timeslide Trigs};

        \node[stat]         (BSN)     [below right of=slides]  {BSN};
        \node[stat]         (BCI)     [below of=BSN]           {BCI};

        \node[actn]         (LRT)     [above right of=BSN]     {Likelihood Ratio};
        \node[rslt]         (signif)  [below right of=LRT]     {Significance};

        \node[data]         (signal)  [above right of=slides]          {Signal Model};
        \node[data]         (noise)   [above right of=LRT]        {Noise Model};

        \path (segs)    edge (omicron) 
              (frames)  edge (omicron) 
              (omicron) edge (sngls)
              (sngls)   edge (coinc)
              (coinc)   edge (0-lag)
                        edge (slides)
              (0-lag)   edge (LIB)
              (slides)  edge (LIB) 
              (LIB)     edge (BSN)
              (LIB)     edge (BCI) 
              (BSN)     edge (LRT)
              (BCI)     edge (LRT)
              (LRT)     edge (signif)
              (signal)  edge (LRT)
              (noise)   edge (LRT) ;

    \end{tikzpicture}
    \caption{A flow chart illustrating the hierarchical structure of the oLIB algorithm.  
        Calibrated strain data and analyzable time segments are fed into Omicron, which produces single-interferometer (IFO) events. 
        The events are down-selected via incoherent clustering, data-quality vetoes, and coincidence.  
        Sets of the most significant analysis (0-lag) and background (timeslide) events are passed onto LIB.  
        The Bayes factors produced by LIB (BSN, BCI) are combined using an LRT.  
        The LRT also requires likelihood models for both the detection (signal) and non-detection (noise) hypotheses.  
        Finally, the LRT provides a measure of each 0-lag event's detection significance.
       }
    \label{Fig.FlowChart}
\end{figure}

In this section, we describe the workflow within oLIB.
The algorithm is graphically depicted in Fig.~\ref{Fig.FlowChart}.
The information-theoretic motivation of the algorithm is provided in \S \ref{SubSec.InformationJustification}, and its implementation is described afterward.
We discuss the individual-detector event generation in \S \ref{SubSec.omicron} and coincidence tests in \S \ref{SubSec.Coincidence}.
\S \ref{SubSec.LIB} describes the LIB analysis and \S \ref{SubSec.LRTDescription} discusses how the LRT is used within oLIB.
Finally, \S \ref{SubSec.NonOptimal} discusses different factors that can cause oLIB's implementation to be sub-optimal.

\subsection{Information-theoretic justification of oLIB's design}\label{SubSec.InformationJustification}

\begin{figure*}
    \begin{tikzpicture}[->,>=stealth', shorten >=1pt, auto, node distance=5.00cm, semithick]
        \tikzstyle{every state}=[rectangle, text=black, rounded corners, fill=none, draw=black]
        
        \node[state] (data)                    {Data} ;
        \node[state] (Bayes)  [right of=data]  {Bayes Factors} ;
        \node[state] (LRT)    [right of=Bayes] {Likelihood Ratio} ;
        \node[state] (Signif) [right of=LRT]   {Significance} ;

        \path (data)  edge node {Compression} node[yshift=-1.5em] {Sufficient Stat} (Bayes)
              (Bayes) edge node {Compression} node[yshift=-1.5em] {Sufficient Stat} (LRT)
              (LRT)   edge node {Extraction}  node[yshift=-1.5em] {LRT} (Signif) ;

    \end{tikzpicture}
    \caption{Schematic of how information is compressed into a significance estimate within the oLIB algorithm.  
	     Ideally, sufficient statistics allow for lossless data compression, and the LRT allows for optimal information extraction.
	    }
    \label{Fig.InfoFlowChart}
\end{figure*}
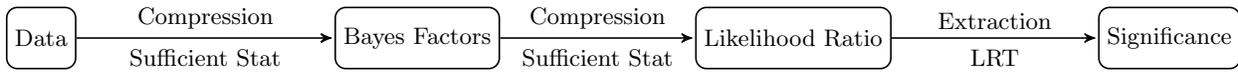

While we have motivated oLIB's design with the idea of preserving information, we have yet to rigorously define this concept.  
Here we provide the framework for an optimal search in an information-theoretic sense.
While other GW-burst search algorithms utilize components of this optimal framework, oLIB is the first to implement it in its entirety.
First, we quantify the qualitative concept of information by utilizing elements from information theory, defining the information in the data stream $\vec{x}$ regarding the signal-versus-noise binary hypothesis $H$ to be their mutual information
\begin{equation}\label{Eq.IdefH}
    I(H;\vec{x}) = \mathcal{H}(H) - \mathcal{H}(H|\vec{x})
\end{equation}
where $\mathcal{H}(H)$ and $\mathcal{H}(H|\vec{x})$ are the entropy and conditional entropy, respectively, of the probability distributions for \Hypo (see Appendix~\ref{App.InfoMultVars} for explicit definitions of entropy).
Because entropy is a measure of distributional uncertainty, the information $I(H;\vec{x})$ quantifies how the uncertainty in the true hypothesis \Hypo is reduced by knowledge of the full data stream $\vec{x}$.

We wish to see how the information changes when we compress the full dimensionality of the data stream $\vec{x}$ into a search statistic $t(\vec{x})$.  
The Data Processing Inequality states that compressing a data vector into a search statistic can only reduce or preserve the amount of accessible information regarding the true hypothesis \Hypo \cite{cover2012elements}:
\begin{equation}\label{Eq.DPI}
    I(H;\vec{x}) \geq I(H;t(\vec{x}))  .
\end{equation}
The Data Processing Inequality becomes an equality for a certain class of statistics known as ``sufficient statistics''.  
A statistic $t(\vec{x})$ is sufficient if and only if it satisfies the relationship
\begin{equation}\label{Eq.SS_def}
    P_{H|x}(H|\vec{x}) = P_{H|t}(H|t(\vec{x})) 
\end{equation}
where each P is a conditional probability distribution, which implies that identical inference of the signal-versus-noise hypothesis can be done with both $\vec{x}$ and $t(\vec{x})$.  

The key design feature in our algorithm is that, for binary hypothesis testing, the likelihood ratio
\begin{equation}
    \Lambda(\vec{x}) \equiv \frac{P_{x|H}(\vec{x}|\text{signal})}{P_{x|H}(\vec{x}|\text{noise})}
\end{equation}
is a sufficient statistic (see Appendix~\ref{App.LRSSProof} for a proof).  
We emphasize that likelihood ratios only compress data losslessly when the likelihoods used in the ratio are the true likelihoods.  
This scenario commonly breaks down in two ways.  
First, the hypothesis used in the likelihood might not be exactly \Hypo.  
Examples include the signal hypothesis referring to a GW being present in the data but with the wrong waveform morphology and the noise hypothesis assuming that the non-Gaussian detector noise is Gaussian.  
Second, the functional form of the probability distributions $P_{x|H}$ might be incorrect even if the hypothesis models are correct.  
In either case, as long as the implemented likelihood ratio is a good approximation to the true one, we expect this information loss to be minimal.

We now explore how to utilize the sufficiency of likelihood-ratio statistics.  By construction, the Bayes factor $B_{i,j}$ computed for any two hypotheses $H_i$ and $H_j$, where
\begin{equation}\label{Eq.BayesFactors}
    B_{i,j} \equiv \frac{P_{x|H}(\vec{x}|H_i)}{P_{x|H}(\vec{x}|H_j)}  ,
\end{equation}
is a likelihood ratio and, in turn, a sufficient statistic.  
Thus, compressing the data vector into a Bayes factor is lossless as long the two hypotheses perfectly describe all possible data realizations.  
Nevertheless, there might be multiple model classes for both signals and noise within the broader signal-versus-noise hypotheses.  
Expanding on our previous examples, GW burst signals can have varying morphologies, and the detector noise may behave as either Gaussian or non-Gaussian noise at different times \cite{S6DetChar}.
We can compute Bayes factors for each of these model class hypotheses, but then we need a way of combining the Bayes factors without losing information about the overall signal-versus-noise hypothesis.  
If we treat the set of Bayes factors for each model hypothesis as another data vector $\vec{x}_B$, then further compression of the data into the likelihood ratio     
\begin{equation}\label{Eq.LRT_Bayes}
    \Lambda(\vec{x}_B) \equiv \frac{P_{x_B|H}(\vec{x}_B|\text{signal})}{P_{x_B|H}(\vec{x}_B|\text{noise})}
\end{equation}
is lossless (see Fig.~\ref{Fig.InfoFlowChart}).  
We prove in Appendix~\ref{App.LRSSProof} that the likelihood ratio for the set of all possible Bayes factors $\Lambda(\vec{x}_B)$ is indeed a sufficient statistic.
\textit{This result is important} because it allows us to construct a single optimal search statistic for an arbitrary number of models.

There is still the question of what happens when no model hypothesis perfectly describes the true signal-versus-noise hypothesis.  
If this is the case, the compression must be lossy.  
It isn't immediately clear what happens if we combine lossy search statistics.  
Fortunately, we show (see Appendix~\ref{App.InfoMultVars}) that adding \textit{any} additional data point $y_{+}$ into an arbitrary data vector $\vec{y}$ can only increase the information contained about the hypothesis H:
\begin{equation}\label{Eq.AddParam}
    I(H;\vec{y},y_{+}) \geq I(H;\vec{y})  .
\end{equation}
Thus, we can combine lossy search statistics with lossless search statistics without losing information, and we can \textit{losslessly compress the information that is contained within lossy search statistics}, both by means of a likelihood ratio.  
We stress that even though information might have been lost in compressing data from $\vec{x}$ to $\vec{x}_B$, further compression of $\vec{x}_B$ can still be lossless.

It should be noted that, to this point, we have only discussed minimizing the loss of information when compressing data. 
However, all of this lossless compression is useless if we do not have an optimal way of extracting the information from the compressed data.  
Just having a compressed statistic ($\Lambda$) containing the maximal amount of information about a model (\Hypo) does not guarantee that any arbitrary estimator $\hat{H}(\Lambda)$ will be optimal.
Fortunately, the Neyman-Pearson lemma \cite{NeymanPearson} argues that a likelihood-ratio test (LRT) maximizes the probability of detection at a given false-alarm probability, so it is an optimal means of information extraction.  
As we will see in \S~\ref{SubSec.LRTDescription}, once we have a likelihood ratio, evaluating an LRT is straightforward.

The implementation of this information-theoretically optimal scheme in oLIB is as follows:
\begin{enumerate}
  \item{Use Omicron to flag stretches of the detector's data streams that contain excess power, which will serve as ``events'' in our further analysis.} 
  \item{For each event, use LIB to calculate Bayes factors across all signal and noise model classes.  
        If the set of signal and noise model classes perfectly describes every data realization, then the compression is lossless.  
        If not, information loss is introduced.
       }
  \item{For each event, use a likelihood ratio $\Lambda$ to combine the information contained within all of the models' Bayes factors.  
        As long as the signal and noise likelihoods used to compute the likelihood ratio are the true likelihoods for each model class, the compression from a set of Bayes factors to $\Lambda$ is lossless.
       }
  \item{Extract the information contained within $\Lambda$ regarding the model \Hypo by using an LRT to map $\Lambda$ into a significance statement.}
\end{enumerate}

In the following sections, we describe these steps in greater detail. 
  
\subsection{Omicron}\label{SubSec.omicron}

Omicron provides fast and accurate identification of statistically significant deviations from Gaussian noise in a single interferometer's data stream.
It is based on the Q transform, which varies the duration of data used within a Fourier transform to maintain a constant quality factor $Q\propto \tau \cdot f_0$, meaning the duration $\tau$ is inversely proportional to the targeted frequency $f_0$ \cite{Qtransform,Gabor,Brown,Omicron}.
By repeatedly decomposing a data stream into several planes of constant Q, Omicron can search for excess power with different characteristic aspect ratios in the time-frequency plane.
In effect, the Q transform is similar to matched-filtering with a bank of sine-Gaussian waveforms, each of which has a characteristic shape in the time-frequency plane and is well-localized.
In this way, oLIB uses Omicron to flag interesting stretches of data and later uses the results of the Omicron analysis to perform all of the down-selection.

\subsection{Coincidence}\label{SubSec.Coincidence}

As mentioned, Omicron matched-filters a bank of sine-Gaussian templates with the entire stretch of data.  
Any template that has a signal-to-noise ratio (SNR) greater-than-or-equal-to a threshold value is recorded as an event.  
However, the presence of excess power alone does not provide strong evidence of a GW because it can also be generated by loud, non-Gaussian noise fluctuations resulting from instrumental or environmental causes. 
We will refer to these fluctuations as noise ``glitches''.  
Furthermore, many of the events are redundant because any excess of power in the data stream can have significant overlap with multiple sine-Gaussian templates, so there are routinely multiple events of different $f_0$ and Q recorded at nearly identical times in the data stream.
For computational reasons, we are motivated to only follow-up the most ``GW-like'' events with LIB.

The strategy of this down-selection naturally falls out of how we define ``GW-like''.  
Even though the oLIB algorithm is designed to detect unmodeled GW bursts, we hypothesize that detectable burst signals exhibit several qualities.  
For example, we expect a GW to leave a specific signature in the data streams of all detectors.  
More precisely, Omicron models these signatures as single sine-Gaussians, so we hypothesize that the events produced by a single burst-like GW will cover similar ranges of $f_0$ and Q in each detector.  
In addition, from General Relativity, we expect GWs to travel at the speed of light, meaning there is a given time window, defined by the physical separations of the detectors, in which GWs can leave this signature.  
As a result, we choose to pass to LIB only the Omicron events whose $f_0$ and Q values are identical across all detectors and whose detection times are consistent with this time-of-flight time window\footnote{We note that requiring \textit{exact} $f_0$ and Q match instead of \textit{close} $f_0$ and Q match may result in a loss of some quieter signals or broadband signals whose SNRs are diluted across large areas of the time-frequency plane.}. 

With this definition of ``GW-like'' in mind, our exact down-selection takes the following form.
First, for each individual detector, we form ``clusters'' of Omicron events with identical $f_0$ and Q that are spaced closely in time.  
More precisely, we open an acceptance gate at the detection time of an event with given $f_0$ and Q, and we leave the gate open so long as an event of identical $f_0$ and Q is found within the time interval $\Delta t_{cluster}$, closing it otherwise.  
Each continuous stretch of acceptance is defined as a ``cluster'' for a template, and we down-select each cluster to the loudest SNR event contained within it.  
The $\Delta t_{cluster}$ used in our analyses is 100 ms, and the analysis results are relatively invariant for time windows of similar size.

The search algorithm also has provisions for the inclusion of data-quality and veto flags as established by the LIGO and Virgo collaborations \cite{S6DetChar,OVL,hVeto}. 
These vetoes are informed by the detectors' auxiliary channels in order to determine whether the excess power in the detectors is caused by environmental, instrumental, or other non-astrophysical sources of noise instead of a flux of GWs.  
To be consistent with the transient searches published by the LIGO-Virgo collaborations, we remove all clustered events that fall within these vetoed times.

Next, we take the set of surviving clustered events and apply a coincidence restriction among the detectors for events of identical $f_0$ and Q.  
More precisely, we only keep clustered events of the $i^{th}$ detector that have a clustered-event counterpart of identical $f_0$ and Q in the $j^{th}$ detector, requiring the corresponding times to fall within a time window $\Delta t_{coin,ij}$.  
For the Hanford and Livingston detectors, the time-of-flight coincidence window we need to apply is about 10 ms.
For each of these coincident events, we can also place thresholds on either the single-detector SNRs $\left( \rho_i \right)$ or the network SNR $\left( \rho_{net}^2 = \sum_{i\in\mathrm{Detectors}} \rho_i^2 \right)$.

Finally, we cluster this set of coincident events one last time so there is at most one event per LIB event window, $\Delta t_{\text{LIB}}$, thereby avoiding redundant LIB runs.  
We do this by iteratively keeping the loudest SNR event in a set of non-overlapping intervals of length $\Delta t_{\text{LIB}}$ until all LIB events are separated by at least $\Delta t_{\text{LIB}}$.  
This set of LIB-clustered coincident events is passed onto LIB for analysis.

\subsection{LIB}\label{SubSec.LIB}

LALInferenceBurst is based on LALInference \cite{LALInference}, a Bayesian parameter estimation and model selection algorithm.
While LALInference assumes that the model waveform is produced by a CBC system (any pairwise combination of a neutron star and a black hole), LIB models short duration signals with ad-hoc waveforms as sine-Gaussians, Gaussians and damped sinusoids.

The standard configuration of oLIB uses LIB with sine-Gaussian templates. 
These templates depend on 9 parameters, which we refer to as $\vec{x}$: central frequency $\left( f_0 \right)$, quality factor (Q), amplitude (the parameter actually used is the $h_\mathrm{rss}$, see \cite{BF2Y}), time, phase, sky position, polarization ellipticity, and orientation of the polarization ellipse.

LIB uses the nested sampling algorithm \cite{NestedSampling} to efficiently sample the 9-D parameter space. 
$N$ ``live points'' are evolved by sampling the prior distribution in order to calculate the Bayesian evidence $P_{x|H}(\vec{x}|H)$ for the data stream $\vec{x}$ and hypothesis H. 
For oLIB, we use the default termination condition \cite{LALInference_nest} that the extra Bayesian evidence one would lose if all of the live points had a likelihood equal to the maximum-likelihood point found is smaller than 0.1.

As shown in Eq.~\ref{Eq.BayesFactors}, the evidences calculated by LIB can be used to construct two Bayes factors.
This first compares a signal model (a sine-Gaussian GW is present in the data of all detectors) to a Gaussian-noise model (only Gaussian noise is present in the data), and we refer to its natural logarithm as the BSN\footnote{The name BSN refers to the comparison of a signal model (S) to a Gaussian-noise model (N).}.
Another compares the same signal model to a noise-glitch model (uncorrelated sine-Gaussian glitches of non-GW origin are present in each instrument), and we refer to its natural logarithm as the BCI\footnote{The name BCI refers to the comparison of a ``coherent'', i.e. correlated, signal model (C) to an ``incoherent'', i.e. uncorrelated, signal model (I).}. 
While the reader is directed to \cite{LALInference,LALInference_nest} for more details about nested sampling and these Bayes factors, we will note that a large BSN implies a loud signal, while a large BCI implies a signal that is highly correlated among the detectors.
As a by-product, LIB produces posterior distributions for all 9 parameters on which the model sine-Gaussian waveform depends. 
While some of them might not be of immediate use since the GW signal may not necessarily be well-matched by a simple sine-Gaussian, it has been shown that the sky position of the source, as measured by LIB, can be used for electromagnetic follow-up \cite{BF2Y}.

\subsection{Likelihood-Ratio Test}\label{SubSec.LRTDescription}

We now explain how we use an LRT to extract information from our search statistics and how to ``train'' this LRT.

\subsubsection{Using the Likelihood-Ratio Test for Detection}\label{SubSubSec.LRTDetection}

The primary purpose of oLIB is to optimally extract the information contained within the data regarding the signal-versus-noise hypothesis and to use this information to make a detection statement.  
As we argue in \S~\ref{SubSec.InformationJustification}, our working assumption is that the Bayes factors produced by LIB compress the dimensionality of the raw data streams while still preserving a sufficiently large fraction of the original information.    

With any n-dimensional set of compressed search statistics $\vec{x}_B =$ BSN,BCI,..., the problem of optimal information extraction immediately suggests the use of an LRT.  
The motivation for this approach comes from the Neyman-Pearson lemma \cite{NeymanPearson}, which states that the LRT is the optimal method of binary hypothesis testing in that it maximizes the probability of successfully detecting a signal at a given false-alarm probability.  
The exact form of the LRT for the signal-versus-noise binary hypothesis test is
\begin{equation}\label{Eq.LRT}
    \Lambda(\vec{x}_B) \equiv \frac{P_{x_B|H}(\vec{x}_B|\text{signal})}{P_{x_B|H}(\vec{x}_B|\text{noise})} \mathop{\gtreqless}^{\text{signal}}_{\text{noise}} \alpha
\end{equation}
where $\Lambda(\vec{x}_B)$ is the value of the likelihood ratio at a coordinate $\vec{x}_B$, $P_{x_B|H}(\vec{x}_B|\text{signal})$ and $P_{x_B|H}(\vec{x}_B|\text{noise})$ are the likelihood distributions of getting the coordinate point $\vec{x}_B$, and $\alpha$ is a threshold value of the likelihood ratio.  
Thus, if $\Lambda(\vec{x}_B)$ is greater than the threshold, we decide that there is a signal present in the data with a false-alarm rate (FAR) set by $\alpha$.
The procedure for establishing a FAR is addressed later in this section.  
Eq.~\ref{Eq.LRT} essentially uses the likelihoods to divide our search statistic parameter space $\mathcal{X}_B$ into regions of detection and non-detection, with $\alpha$ determining the boundary.  

We emphasize that the LRT allows us to optimally compress the n-dimensional vector of search statistics $\vec{x}_B$ into a \textit{single scalar measure of significance} $\Lambda(\vec{x}_B)$.
While several GW searches attempt to combine information from multiple search statistics \cite{S6Burst,BWcomplexity}, only a few \cite{GSTLALLRT, CosmicStrings,Biswas:2012} utilize the optimality of the LRT.
The LRT has the property that the FAR associated with a decision is a monotonically decreasing function of the threshold $\alpha$.
Thus, the events with the largest values of $\Lambda(\vec{x})$ are necessarily the most significant events encountered.  

This monotonicity allows us to rank events by $\Lambda(\vec{x}_B)$ and lets us empirically estimate the FAR of events.
In order to achieve this, the oLIB algorithm is run end-to-end on a stretch of background data, producing a vector of Bayes factors $\vec{x}_{B,i}$ for each background event $i$.  
We then calculate a value of the likelihood ratio $\Lambda(\vec{x}_{B,i})$ for each background event $i$.
Using the total coincident livetime of our background analysis (i.e., the duration of time during which an event could have been generated), we can approximate the FAR of a threshold $\alpha$ with a simple counting experiment:
\begin{equation}\label{Eq.FAR}
    \begin{split}
        \text{FAR}&(\alpha) \approx \\
        &\frac{\text{No. of background events with } \Lambda(\vec{x}_{B,i}) \geq \alpha}{\text{Total coincident livetime of background analysis}}
    \end{split}
\end{equation}
Finally, for any detection candidate $j$ (i.e., any event generated in the analysis data), the oLIB algorithm produces a vector of search statistics $\vec{x}_{B,j}$.  
By calculating the likelihood ratio $\Lambda(\vec{x}_{B,j})$ and setting $\Lambda(\vec{x}_{B,j}) = \alpha$, we can use Eq.~\ref{Eq.FAR} to estimate the FAR of event $j$.

Background estimation for GW detectors is complicated by the fact that it is impossible to isolate the detectors from any incident GW flux, meaning GWs are always present in the detectors' data streams, even if they reside below the detector noise floor. 
Nevertheless, we can transform a stretch of data into a stretch of data without any ``coherent'' GWs by a technique known as data timesliding \cite{timeslides}.  
To accomplish this, we shift the timestamps of one detector's data stream in bulk (i.e., we apply the same time shift to every discrete time sample) with respect to another detector's data stream before doing coincidence.  
If this timeshift is greater than the time-of-flight between the detectors for a GW, then the GW-induced correlation of the data streams becomes non-astrophysical in our model.  
Thus, any events found in coincidence among the detectors can be modeled as non-Gaussian (commonly Poisson-distributed) noise glitches that occur simultaneously but independently in the detectors.  
In summary, timeslides provide a method for approximating the noise-only background rate of our detectors using real detector data, and as a result, we commonly refer to our analysis data as the ``0-lag'' data\footnote{There are subtleties involved with timeslides with regards to which 0-lag coincidences to remove from the data before doing the timeslides in order to reduce GW contamination.  For this paper, we do not remove any 0-lag coincidences, although it has become a common procedure to remove high-confidence GW detections from the timeslided data.}.

\subsubsection{Training the Likelihood-Ratio Test}\label{SubSubSec.LRTTraining}

We stress again that while this LRT-based method is straightforward and can be considered optimal under several criteria (information preservation and extraction), all optimality statements assume we have access to the true likelihood distributions for both our signal and noise hypotheses.  
Any inaccuracies in our likelihoods will lead to both lossy compression and sub-optimal information extraction.  
Thus, we need to accurately model these likelihood distributions before we estimate the significance of any events.

We need models for both the signal and noise likelihoods, $P_{x_B|H}(\vec{x}_B|\text{signal})$ and $P_{x_B|H}(\vec{x}_B|\text{noise})$, respectively.  
We choose to implement an empirical approach to our modeling in which we simulate large sets of signal and noise events and calculate the vector of Bayes factors $\vec{x}_B$ for each.  
We then fit the resulting distribution of $\vec{x}_B$ using non-parametric regression, specifically the Gaussian kernel density estimation (KDE) described in detail in Appendix~\ref{App.KDE}.
We refer to this process as ``training'' the LRT.

\subsection{Summary of how optimality breaks down in oLIB's implementation}\label{SubSec.NonOptimal}

Although we justified the optimality of oLIB's design in \S~\ref{SubSec.InformationJustification}, such optimality is not acheived in practice.  Here we will briefly review and discuss the ways in which oLIB's implementation can lead to sub-optimal performance.
\begin{enumerate}
  \item{As previously mentioned, oLIB models the GW signals as sine-Gaussians, and the noise as Gaussian with potential sine-Gaussian glitches.  
  If the signals and detector noise only ever take these forms, then oLIB's data compression should lose no information concerning \Hypo.  
  However, in most scenarios, these models are only approximations, so information loss is introduced.  
  Including a wider range of models in our vector of Bayes factors $\vec{x}_B$ could help to suppress this information loss, but as we will see in \S~\ref{SubSubSec.EffMorph}, oLIB can detect a wide-range of morphologies regardless, suggesting that this information loss is not significant.
  }
  \item{Although we are treating GW bursts as unmodeled, in practice we need to enforce a minimal set of assumptions in order to distinguish GW signals from noise.  
  We can obtain populations of noise events through timeslides, but we must make assumptions regarding the population of GW burst signals.  
  These assumptions involve choosing the set of signal morphologies on which to train (e.g., sine-Gaussian signals) and then specifying the distribution of these morphologies' intrinsic parameters (e.g., the distributions of $f_0$ and Q for sine-Gaussian signals).  
  There are also distibutionss for the extrinsic parameters, such as the source sky location, but these distributions can be modeled and justified theoretically (e.g., uniformly in volume when considering distant sources), making our asumptions less arbitrary.
  While the arbitrariness of selecting the signals' intrinsic population may seem like a substantial limitation for oLIB, the impact of training on different population models is actually quite small.  
  We explore this feature explicitly in \S~\ref{SubSubSec.EffTrain}, but the intuitive understanding is as follows:  any GW signal interacts with oLIB in a different manner than accidental noise coincidences, meaning we can train our LRT to distinguish GW signals from incoherent noise regardless of the exact form of our training populations.
  }
  \item{In order to accurately model our LRT likelihood functions non-parametrically, we need a large empirical data set on which to train.
  To be sure, some extent of modeling error will be introduced by having a finite data set, although this error will be negligible if the training set is sufficiently large.
  Furthermore, there is a trade-off between the information gained by adding a search statistic to $\vec{x}_B$ and the accuracy of our likelihood modeling.  
  Although we show that adding a search statistic can only increase the information contained within $\vec{x}_B$, it also increases the dimensionality of the search-statistic parameter space $\mathcal{X}_B$.  
  Increasing the dimensionality of a parameter space further dilutes regions where empirical training points were already sparse, leading to greater modeling errors in the distribution's tail.
  Thus, because our optimality conditions require the use of the true likelihoods, adding a weakly-informative search statistic can harm the performance of our algorithm.
  }
  \item{Finally, the LRT is an optimal decision-making method at the false-alarm probability defined by its threshold $\alpha$.  
  Our estimate of the FAR given by Eq.~\ref{Eq.FAR} is an approximation that approaches the true value in the limit that both the number of background events exceeding the detection threshold and the coincident livetime become infinite.
  If we are estimating the FAR with too few above-threshold background events, our estimate may be poor, leading to sub-optimal performance of the LRT (either in rejecting false-alarms or detecting GW signals) at the claimed FAR.
  }
\end{enumerate}

\section{A Sample Analysis}\label{Sec.Application}

We perform a proof-of-concept simulation in order to demonstrate more illustratively how oLIB functions.  
Performance comparsions with other search algorthims are an integral part of the real GW-burst searches that have been \cite{BCP,O1AllSky} and will be completed in the advanced detector era.  
Completed comparisons have shown that oLIB is competitive with other GW burst search algorithms in terms of sensitivity, and is the most sensitive search algorithm in certain regions of the short-duration GW burst parameter space \cite{O1AllSky}.

In order to illustrate typical features of oLIB's end-to-end performance, we undertook the analysis of three days worth of data from the sixth science run of initial LIGO (S6) \cite{S6Data}.  
Specifically, we ran on science time segments produced for the Hanford (H1) and Livingston (L1) detectors between 14-17 September 2010.
These dates were chosen since they contain a blind chirp-like hardware injection \cite{BigDog} (removed from our analysis time segments).
The science time segments signify that the instruments were in proper states for observation, and additional data-quality vetoes were applied to these time segments.

In order to ensure that we encountered a sufficient number of detection candidates, varying both in significance and morphology, we injected simulated GW waveforms into the data streams.  
These injections were taken from the S6 Burst injection set \cite{S6Burst}, and we injected them at multiple amplitude scale factors to ensure that a large range of SNRs were covered.
We injected three morphologies:  1.) sine-Gaussians (SG), described earlier; 2.) Gaussians (GA), which are characterized by their duration $\tau$; and 3.) white-noise bursts (WNB), which consist of random white noise within a Gaussian envelope and are characterized by a starting frequency $f_0$, a duration $\tau$, and a bandwidth $\Delta f$. 
More detailed information regarding these burst morphologies can be found in \cite{S6Burst} and \cite{BF2Y}.  

\subsection{Coincidence Results}\label{SubSec.Incoherent}

\begin{table*}[htb]
    \centering
    \scriptsize
    \caption{Summary of the event rate at each step in the pre-LIB down-selection.  
	     The numbers given represent the set of events immediately after the quoted down-selection is applied.
	     The post-coincidence events span all of the timeslides, which is responsible for the increase in livetime.
	    }
    \label{Tab.IncoherentStats}
    \begin{tabular}[c]{c||c|c|c}
        \hline
        Step & Number of Events & Total livetime (s) & Trigger Rate (Hz) \\
        \hline\hline
        Unclustered H1 & 1410060 & $1.46\times10^{5}$ & 9.66 \\
        Unclustered L1 & 1786080 & $1.46\times10^{5}$ & 12.2 \\
        \hline
        Clustered H1 & 623016 & $1.46\times10^{5}$ & 4.27 \\
        Clustered L1 & 676208 & $1.46\times10^{5}$ & 4.63 \\
        \hline
        Clustered H1, post-Vetoes & 585700 & $1.45\times10^{5}$ & 4.04 \\
        Clustered L1, post-Vetoes & 630606 & $1.45\times10^{5}$ & 4.35 \\
        \hline
        Coincident H1L1, network SNR $\geq 6.5\sqrt2$ & 32779 & $3.15\times10^{8}$ & $1.04\times10^{-4}$ \\
        \hline
        LIB-clustering H1L1 & 18599 & $3.15\times10^{8}$ & $5.90\times10^{-5}$ \\
    \end{tabular}
\end{table*}

\begin{figure}[h!]
        \scalebox{.85}{\includegraphics[width=0.5\textwidth]{\PlotPath/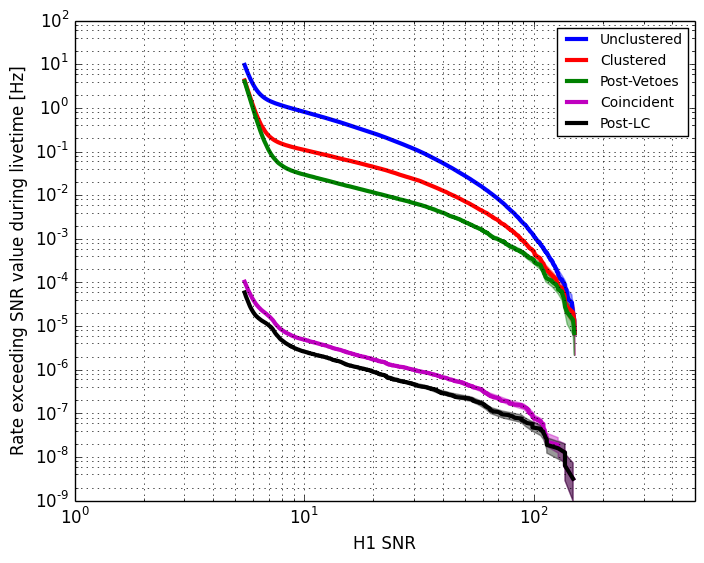}}
        \scalebox{.85}{\includegraphics[width=0.5\textwidth]{\PlotPath/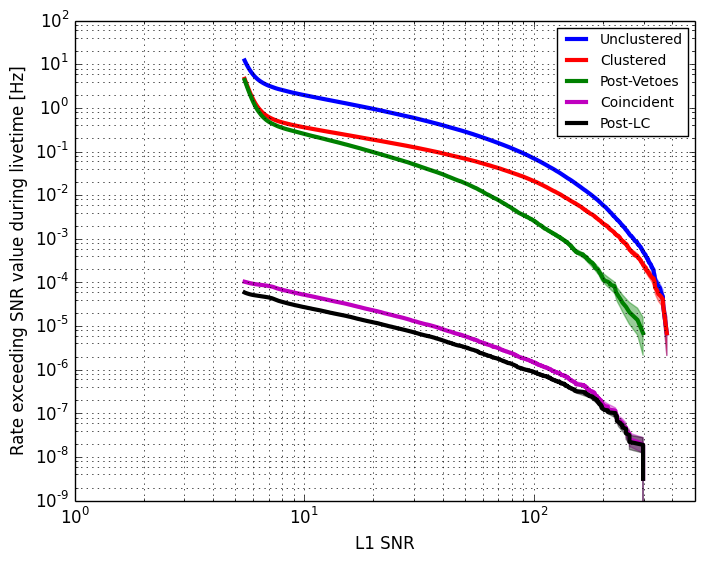}}
        \caption{The rates at which the events generated by Omicron exceeded a given value of SNR in Hanford (top) and Livingston (bottom).  
                 These events are grouped by the down-selection steps they have just survived:  either clustering, data-quality vetoes, timing coincidence, or LIB-window clustering (LC).
                 The 68\% confidence regions shown are derived from a binomial process with a uniform prior on the true rate.
                }
        \label{Fig.TrigRates}
\end{figure}

We ran Omicron separately over both the H1 and L1 data streams, analyzing the frequency band of 64-2048 Hz.  
A single-detector SNR threshold of 5.5 was required for Omicron to identify events.  
Then, using the raw Omicron events, we clustered all identical-template events (as described in \S~\ref{SubSec.Coincidence}) using a clustering window of $\Delta t_{cluster} = 100\text{ms}$.  
We also removed vetoed livetime \cite{S6Burst} from our analysis at this clustering step.
Next, we performed identical-template coincidence between the detectors using the coincidence window $\Delta t_{coin,H1L1} = 10$ ms and required the network SNR to be greater than $6.5\sqrt{2}$.  
We shifted the L1 injection data stream with respect to the H1 injection data stream 2500 times, from -1250s to 1250s in 1s increments, in order to estimate the background.  
Finally, this final set of coincident events was clustered so that only one event was present per LIB event time-window of $\Delta t_{\text{LIB}} = 100$ ms.  

The net result of our down-selection is illustrated in Table~\ref{Tab.IncoherentStats} and Fig.~\ref{Fig.TrigRates}. 
Table~\ref{Tab.IncoherentStats} shows the total number of events, total livetime analyzed, and the total event rate at each step of the down-selection.  
Fig.~\ref{Fig.TrigRates} shows the rate at which events exceeding a given SNR occur in each step of the incoherent analysis for H1 and L1, respectively.  
From this data, we see that the clustering reduced the event rate by roughly a factor of 2-3, and, as expected, most of the discarded events were low-SNR events that were clustered into high-SNR events.  
The application of data vetoes reduced the event rate by less than 10\%, and removed low-to-medium SNR events for H1 and medium-to-high SNR events for L1.

The constraint of identical-template timing coincidence was responsible for our most significant reduction in event rate, lowering the total rate by $\sim$ 5 orders-of-magnitude.  
This fractional reduction appears to be roughly constant, within errors, across all SNRs, which is consistent with a simple Poisson coincidence model.  
Finally, the LIB clustering reduced the event rate by a factor of up to 2, which, characteristic of clustering, discarded low-SNR events when they were clustered into high-SNR events.  
In summary, this pre-LIB down-selection reduced the raw Omicron event rate by $\sim$ 6 orders-of-magnitude.

\subsection{LIB Results}\label{SubSec.Coherent}

We ran LIB over all events surviving the down selection, both for the injection-filled and the injection-free background data sets.  
Our LIB runs used 256 live points and completed 256 MCMC jumps when generating new coordinates for the live points \cite{LALInference_nest}.  
Our sampling frequency was 4096 Hz, and our priors were set to be uniform between 64 Hz and 2048 Hz for $f_0$ and uniform between 2 and 110 for Q.  
For both the signal and noise-glitch models, we assumed sky location and signal-strength priors consistent with a uniform-in-volume distribution.  
This can be justified astrophysically for our signal model; however, it is less justifiable for the noise-glitch model.  
Ongoing investigations are studying the distributions of apparent sky position and $h_\mathrm{rss}$ for the noise-glitch model, but using the uniform-in-volume prior is a conservative approach since it biases the noise and signal model towards each other\footnote{To be sure, these biases are negligible for the likelihood-dominated inference of loud signals.}.

Because the calculation of the Bayes factors requires LIB to integrate over the entirety of the sine-Gaussian parameter space, it is the most computationally expensive step in the oLIB algorithm.  
Executing the Omicron, coincidence, and likelihood-ratio steps all take place on the timescales of a few tens of seconds.  
When run on a single 3 GHz CPU core, the joint-detector (H1L1) LIB analyses had a mean runtime of about 1100 s, while the single detector LIB analyses had mean runtimes of about 200 s and 600 s for Hanford and Livingston, respectively.
Signal-like events take longer to analyze because they have more concentrated likelihood distributions than noise-like events, which LIB needs more iterations to integrate over accurately.  
That the average runtimes were longer for L1 than for H1 is consistent with the analysis of \ref{SubSec.Incoherent}, which shows that L1 contained a greater number of high-SNR events than H1.  
Finally, LIB took longer to run jointly over both detector's data streams than it did to run over each detector's data streams individually because the joint-likelihood constraint more strongly distinguishes signals from noise than the single-detector likelihoods\cite{LALInference_nest}.
It should be noted that the limiting LIB timescale for GW signals is similarly a few thousands of seconds, or tens of minutes.

\subsection{LRT Results}\label{SubSec.LRTResults}

\begin{figure}[h!]
    \scalebox{.85}{\includegraphics[width=0.5\textwidth]{\PlotPath/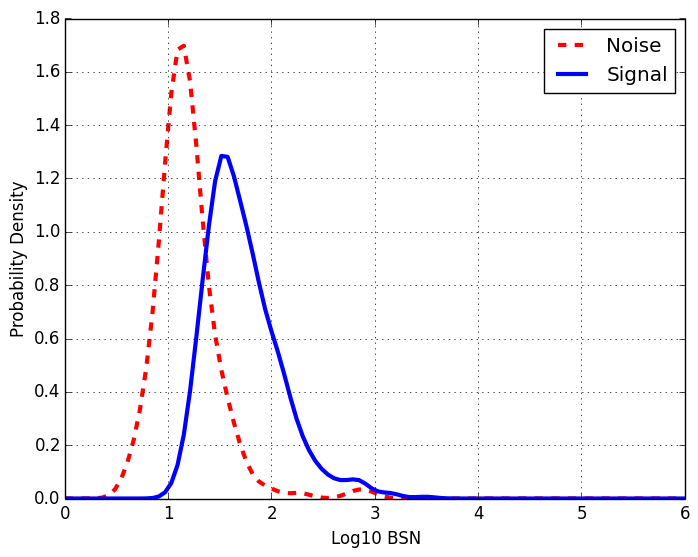}}
    \scalebox{.85}{\includegraphics[width=0.5\textwidth]{\PlotPath/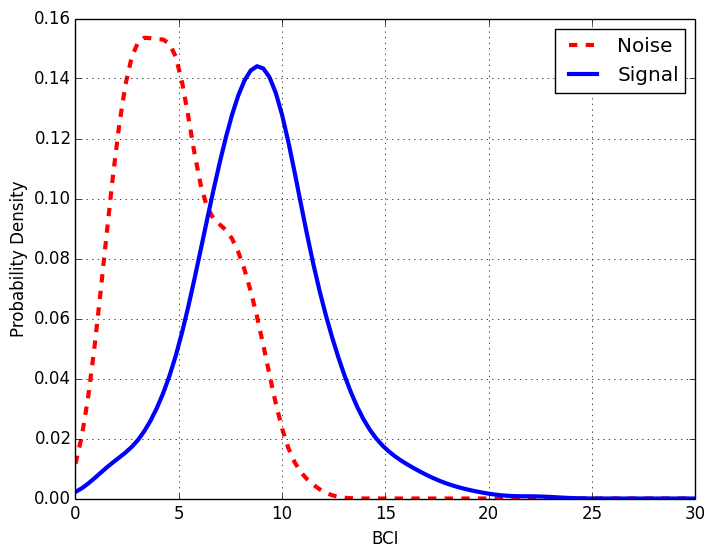}}
    \caption{The 1-dimensional likelihoods for each Bayes factor.  
             In this figure, the signal training population consisted of both sine-Gaussian and white-noise burst signals.  
             The likelihood ratio $\Lambda$ is found by taking the ratio of the signal and noise likelihoods.
            }
    \label{Fig.Likelihoods_1d}
\end{figure}

\begin{figure}[h!]
    \scalebox{.85}{\includegraphics[width=0.5\textwidth]{\PlotPath/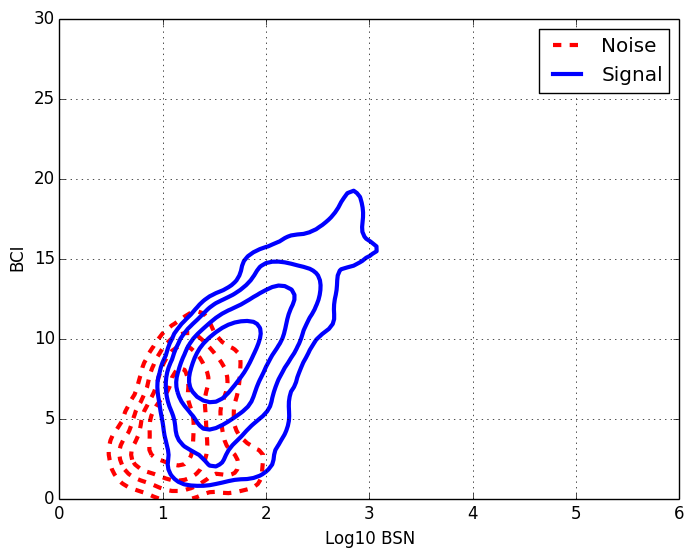}}
    \caption{The 2-dimensional likelihoods for the Bayes factors.  
	     The contours shown correspond to the 0.5-sigma, 1-sigma, 1.5-sigma, and 2-sigma central confidence regions.
	     The likelihood ratio $\Lambda$ is found by taking the ratio of the signal and noise likelihoods.
            }
    \label{Fig.Likelihoods_2d}
\end{figure}

We trained our likelihoods using Gaussian KDE optimized by the Kullback-Leibler distance minimization criterion described in Appendix~\ref{App.KDE}.
Because the values of the BSN covered a large dynamical range, we actually trained on log$_{10}$ BSN to improve performance.
Also, because the Bayes factors are constructed so that positive values of their logarithm favor the signal model over the noise model, we placed an exclusion cut on all events with a BSN or BCI less than 1 (with 1 being chosen instead of 0 because we take the logarithm of BSN).
Finally we also placed an exclusion cut on all events with a BSN or BCI greater than $10^6$ in order to remove events with extremely large, non-astrophysical SNRs that are characteristic of some morphologies of noise glitches.
We trained our noise likelihoods using 100 non-background timeslides.  
We trained our signal likelihoods on a set of astrophysically distributed SGs and WNBs, the exact populations of which are described in \S~\ref{SubSubSec.EffTrain}.  
Examples of the resulting 1-D and 2-D likelihood distributions are shown in Figs.~\ref{Fig.Likelihoods_1d} and ~\ref{Fig.Likelihoods_2d}, respectively.  
The distributions of the Bayes factors follow the general behavior we expect from them by construction:  both BCI and BSN have more support at higher values for signals than for noise.

These distributions illustrate how information is gained by using a combination of search statistics.  
For example, referencing the BCI-BSN plot in Fig.~\ref{Fig.Likelihoods_2d}, we see that the outermost contour of the noise distribution is completely rejected by classifying any event with a BCI below 12 as noise.  
However, we can remove the same noise contour while retaining more of the signal distribution by classifying as noise any event with a BCI below 12 \textit{and} a log10 BSN below 2 as noise.  
Effectively, we constructed a more powerful decision surface in the latter case.  
The LRT optimally constructs this decision surface, thus maximizing the probability of detecting a signal at a given false alarm probability.  
Furthermore, the amount of information contained within the search statistics defines how well the noise and signal distributions can be separated, which in turn determines how powerful the optimal decision surface is in terms of distinguishing signal from noise.

\begin{figure}[h!]
    \scalebox{.85}{\includegraphics[width=0.5\textwidth]{\PlotPath/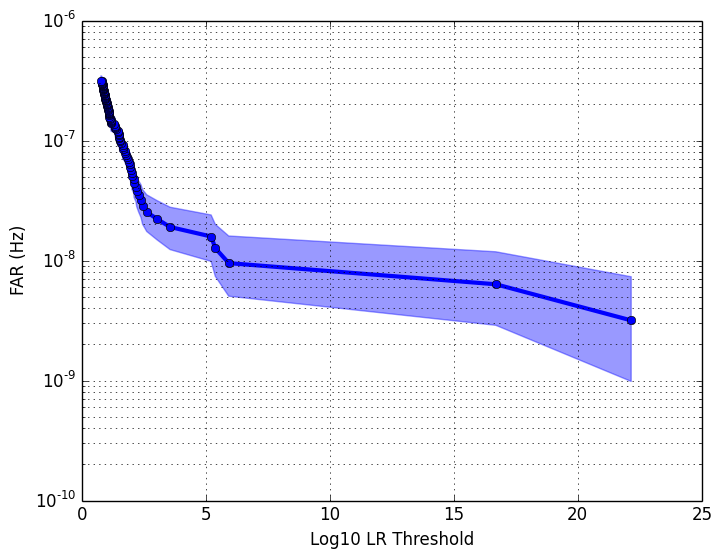}}
    \caption{The FAR achieved by setting a given likelihood-ratio threshold for detection.  
             The LRT shown here is trained on both sine-Gaussian and white-noise burst signal populations and is evaluated using BCI and BSN as search statistics.    
             The 68\% confidence regions shown are derived from a binomial process with a uniform prior on the true rate.
             }
    \label{Fig.LLRDistribution}
\end{figure}

With these estimated likelihoods in hand, we were able to use our background data to estimate the FAR assigned to events of various $\Lambda$, which is shown in Fig.~\ref{Fig.LLRDistribution}. 
We will now explore how the detection efficiency of oLIB varies as a function of:  1) the injected waveform morphologies, 2) the combination of Bayes factors used in the LRT, and 3) the signal populations used to train the LRT.

\subsubsection{Efficiency vs. Signal Morphology}\label{SubSubSec.EffMorph}

\begin{figure}[h!]
    \scalebox{.85}{\includegraphics[width=0.5\textwidth]{\PlotPath/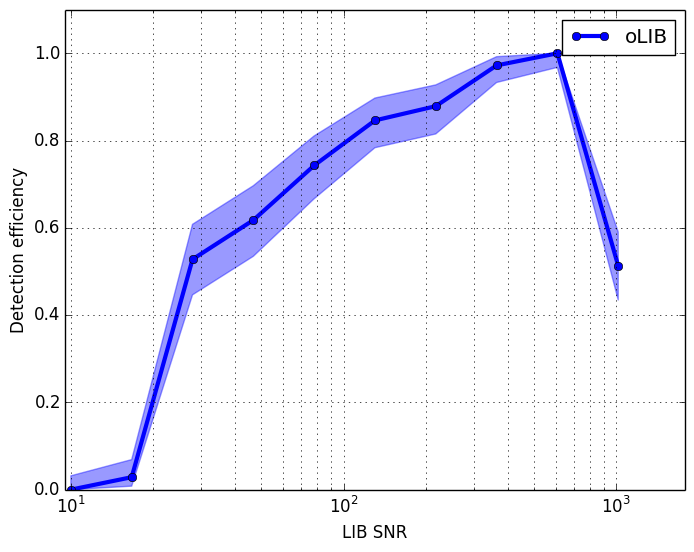}}
    \scalebox{.85}{\includegraphics[width=0.5\textwidth]{\PlotPath/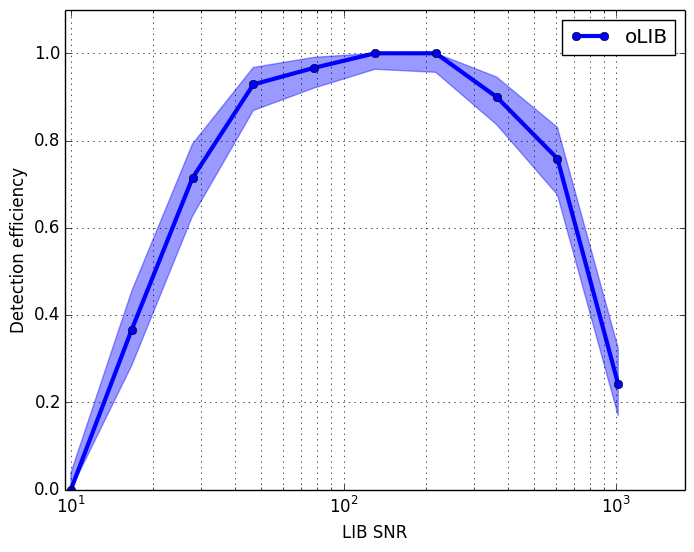}}
    \scalebox{.85}{\includegraphics[width=0.5\textwidth]{\PlotPath/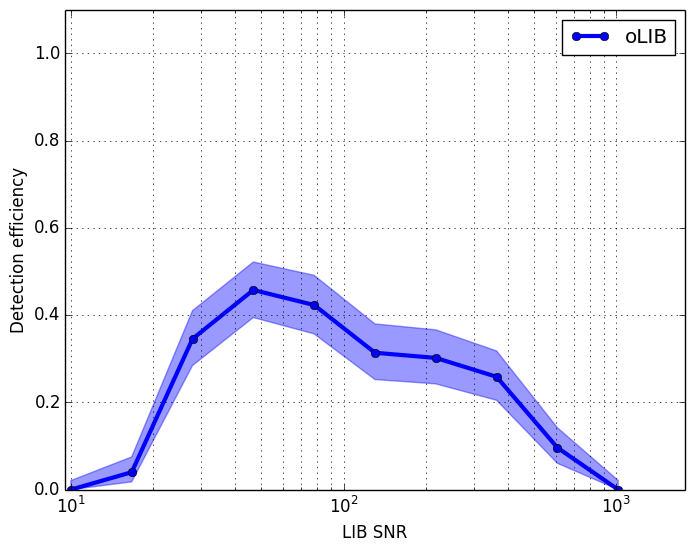}}
    \caption{The detection efficiency as a function of injected network SNR (as estimated by LIB) at a FAR of 1 per decade for three different morphologies of injected waveforms: sine-Gaussian waveforms (top) with $f_0 = 153$ Hz and $Q = 8.9$, Gaussian waveforms (middle) with $\tau = 2.5$ ms , and white-noise burst waveforms (bottom) with $f_0 = 1000$ Hz, $\Delta = 10$ Hz, and $\tau = 100$ ms.  
             The LRT used here is trained on both sine-Gaussians and white-noise bursts and is evaluated using both the BCI and the BSN as search statistics.  
             The 68\% confidence region shown is derived from a binomial process with a uniform prior on the true detection efficiency.}
    \label{Fig.Eff_v_Morph}
\end{figure}

\begin{table*}[htb]
    \centering
    \scriptsize
    \caption{The SNRs at which the detection efficiency reached 10\%, 50\%, and 90\% for different injected signal morphologies using LRTs corresponding to several different FARs.
	     The LRTs were trained on both SG and WNB signals and were evaluated using BCI and BSN as search statistics.  
	     Detection efficiencies that were never reached at any SNR are denoted as N/A.
	     }
    \label{Tab.Eff_v_Morph}
    \begin{tabular}[c]{c||c|c|c||c|c|c||c|c|c}
        \hline
        \multicolumn{1}{c||}{} & \multicolumn{3}{c||}{FAR: 1 per decade ($3\times10^{-9}$ Hz)} & \multicolumn{3}{c||}{FAR: 1 per year ($3\times10^{-8}$ Hz)} & \multicolumn{3}{c}{FAR: 1 per month ($3\times10^{-7}$ Hz)} \\
        \hline
        Morphology & SNR$_{10\%}$ & SNR$_{50\%}$ & SNR$_{90\%}$ & SNR$_{10\%}$ & SNR$_{50\%}$ & SNR$_{90\%}$ & SNR$_{10\%}$ & SNR$_{50\%}$ & SNR$_{90\%}$ \\
        \hline\hline
        SG: $f_0=100$ Hz, $Q=8.9$ & 16 & 23 & 130 & $<$10 & 12 & 80 & $<$10 & $<$10 & 80 \\
        SG: $f_0=153$ Hz, $Q=8.9$ & 18 & 27 & 240 & $<$10 & 13 & 45 & $<$10 & $<$10 & 40 \\
        SG: $f_0=1053$ Hz, $Q=9$ & 13 & 22 & 72 & $<$10 & 11 & 52 & $<$10 & $<$10 & 17 \\
        \hline
        GA: $\tau=0.1$ ms & 14 & 27 & N/A & $<$10 & 13 & 68 & $<$10 & 12 & 54 \\
        GA: $\tau=2.5$ ms & 12 & 20 & 44 & $<$10 & $<$10 & 29 & $<$10 & $<$10 & 15 \\
        GA: $\tau=4.0$ ms & 12 & 23 & 73 & $<$10 & 11 & 56 & $<$10 & $<$10 & 56 \\
        \hline
        WNB: $f_0=100$ Hz, $\Delta f=100$ Hz, $\tau=100$ ms & 35 & N/A & N/A & 15 & N/A & N/A & 11 & N/A & N/A \\
        WNB: $f_0=250$ Hz, $\Delta f=100$ Hz, $\tau=100$ ms & 37 & N/A & N/A & 16 & N/A & N/A & 11 & N/A & N/A \\
        WNB: $f_0=1000$ Hz, $\Delta f=10$ Hz, $\tau=100$ ms & 18 & N/A & N/A & 11 & 26 & N/A & 11 & 23 & N/A \\
        WNB: $f_0=1000$ Hz, $\Delta f=1000$ Hz, $\tau=10$ ms & 31 & N/A & N/A & 13 & 27 & N/A & 11 & 24 & N/A \\
    \end{tabular}
\end{table*}

Here we examine how oLIB's detection efficiency changes as a function of the injected GW waveform morphology.  
As noted, LIB uses sine-Gaussian templates when calculating the Bayes Factors, and thus we expect the oLIB algorithm to best recover sine-Gaussian signals.  
Fig.~\ref{Fig.Eff_v_Morph} shows the detection efficiency for several different injected morphologies as a function of the signal SNR residing within LIB's frequency bandwidth.  
We note that, because Gaussian signals are centered at a frequency of zero, only a fraction of their total SNR is accessible to LIB.  
The shapes of the particular curves shown here are characteristic of the different morphologies in general.  
As expected, the efficiency curves for sine-Gaussian and Gaussian (which are sine-Gaussians in the limit of $f_0 \rightarrow 0$ and $Q \rightarrow 0$) morphologies rise to unity before falling off at large SNRs that are non-astrophysical as a result of our exclusion cut on large-BSN events.
The efficiency curves for white-noise bursts rise similarly to those of sine-Gaussians and Gaussians for low SNRs, but fall off before ever reaching unity.

This behavior is understandable when considering the residuals of template mismatch.  
At low SNRs, the mismatch between the data stream and signal template is dominated by noise since the noise amplitude is comparable to the signal-template mismatch.  
As the SNR of the signal increases, the amplitude of the noise remains the same, but the amplitude of the the signal-template mismatch residuals grows linearly with the signal amplitude.  
Thus, if a template cannot perfectly match the form of a GW signal, the signal-template mismatch will dominate the noise-template mismatch in the limit of high SNRs.  
In practice, these large residuals cause the BCI to become extremely negative for high-SNR white-noise burst signals, which causes the LRT to declare them noise glitches despite having large BSN.   
While this behavior is unfortunate, we expect these types of loud-SNR signals to be extremely rare. 
For example, GW150914 is considered to be a high-SNR, non-sine-Gaussian event with its SNR of 24 \cite{GW150914}, and it was detected confidently by oLIB \cite{BCP}.

Table~\ref{Tab.Eff_v_Morph} shows more extensive results of our simulations.  
The results span three different LRTs, each using a detection threshold corresponding to a different FAR to give a rough picture of how detection efficiency scales with FAR. 
We emphasize that the FAR is better estimated at higher values since there are more background events above threshold at these values.
oLIB efficiently detects sine-Gaussians and Gaussians of varying morphology roughly equally-well at all three FARs.
It performs roughly a factor of 2 worse for white-noise burst injections of varying morphology.

\subsubsection{Efficiency vs. LRT Parameters}\label{SubSubSec.EffParams}

\begin{table*}[htb]
    \centering
    \scriptsize
    \caption{The SNRs at which the detection efficiency reached 10\%, 50\%, and 90\% for different injected signal morphologies using LRTs evaluated with several different vectors of Bayes factors.
	     The LRTs were trained on both SG and WNB signals and corresponded to an FAR of 1 per year.  
	     Detection efficiencies that were never reached at any SNR are denoted as N/A.
	     }
    \label{Tab.Eff_v_Params}
    \begin{tabular}[c]{c||c|c|c||c|c|c||c|c|c}
        \hline
        \multicolumn{1}{c||}{} & \multicolumn{3}{c||}{BSN} & \multicolumn{3}{c||}{BCI} & \multicolumn{3}{c}{BSN-BCI} \\
        \hline
        Morphology & SNR$_{10\%}$ & SNR$_{50\%}$ & SNR$_{90\%}$ & SNR$_{10\%}$ & SNR$_{50\%}$ & SNR$_{90\%}$ & SNR$_{10\%}$ & SNR$_{50\%}$ & SNR$_{90\%}$ \\
        \hline\hline
        SG: $f_0=100$ Hz, $Q=8.9$ & 12 & 49 & 75 & 12 & 20 & 160 & $<$10 & 12 & 80 \\
        SG: $f_0=153$ Hz, $Q=8.9$ & 13 & 45 & 77 & 17 & 26 & 310 & $<$10 & 13 & 45 \\
        SG: $f_0=1053$ Hz, $Q=9$ & 15 & 42 & 71 & $<$10 & 17 & 110 & $<$10 & 11 & 52 \\
        \hline
        GA: $\tau=0.1$ ms & 28 & 51 & 72 & $<$10 & 21 & N/A & $<$10 & 13 & 68 \\
        GA: $\tau=2.5$ ms & 14 & 43 & 68 & $<$10 & 14 & 44 & $<$10 & $<$10 & 29 \\
        GA: $\tau=4.0$ ms & 31 & 45 & 73 & $<$10 & 13 & 73 & $<$10 & 11 & 56 \\
        \hline
        WNB: $f_0=100$ Hz, $\Delta f=100$ Hz, $\tau=100$ ms & 56 & N/A & N/A & 29 & N/A & N/A & 15 & N/A & N/A \\
        WNB: $f_0=250$ Hz, $\Delta f=100$ Hz, $\tau=100$ ms & 57 & N/A & N/A & 33 & N/A & N/A & 16 & N/A & N/A \\
        WNB: $f_0=1000$ Hz, $\Delta f=10$ Hz, $\tau=100$ ms & 50 & 74 & N/A & 12 & N/A & N/A & 11 & 26 & N/A \\
        WNB: $f_0=1000$ Hz, $\Delta f=1000$ Hz, $\tau=10$ ms & 60 & N/A & N/A & 15 & N/A & N/A & 13 & 27 & N/A \\
    \end{tabular}
\end{table*}

We now explore how the detection efficiency varies as a function of the search statistics used as parameters in our LRT.  
As shown in Appendix~\ref{App.LRTOptimal}, likelihood ratios are sufficient statistics that optimally preserve the information contained within a set of search statistics about the binary signal-versus-noise hypothesis and adding another search statistic to the analysis can only increase the information.  
Thus, we would expect that if the likelihoods used in our LRT were accurate, an LRT with a greater number of search statistics would have a better-than-or-equal signal detection efficiency than an LRT utilizing fewer search statistics. 
We consider three different LRTs:  one where the BSN is the only search statistic, one where the BCI is the only search statistic, and one where both the BCI and BSN are used as search statistics.

Table~\ref{Tab.Eff_v_Params} characterizes the detection efficiency for each of these LRTs.
In order to ensure we have a reasonably accurate estimate of the FAR, we compare the efficiencies at a FAR of 1 per year.
As expected, the BCI-BSN LRT outperforms both the BSN-only and the BCI-only LRT across all morphologies 
We also note that the BCI-only LRT outperforms the BSN-only LRT, meaning it is the more informative Bayes factor for detection in real detector noise.  

\subsubsection{Efficiency vs. Training Population}\label{SubSubSec.EffTrain}

\begin{table*}[htb]
    \centering
    \scriptsize
    \caption{The SNRs at which the detection efficiency reached 10\%, 50\%, and 90\% for different injected signal morphologies using LRTs trained on several different signal populations.
	     The LRTs were evaluated using BCI and BSN as search statistics and corresponded to an FAR of 1 per year.  
	     Detection efficiencies that were never reached at any SNR are denoted as N/A.
	     }
    \label{Tab.Eff_v_Training}
    \begin{tabular}[c]{c||c|c|c||c|c|c||c|c|c}
        \hline
        \multicolumn{1}{c||}{} & \multicolumn{3}{c||}{SG} & \multicolumn{3}{c||}{WNB} & \multicolumn{3}{c}{SG and WNB} \\
        \hline
        Morphology & SNR$_{10\%}$ & SNR$_{50\%}$ & SNR$_{90\%}$ & SNR$_{10\%}$ & SNR$_{50\%}$ & SNR$_{90\%}$ & SNR$_{10\%}$ & SNR$_{50\%}$ & SNR$_{90\%}$ \\
        \hline\hline
        SG: $f_0=100$ Hz, $Q=8.9$ & $<$10 & 12 & 130 & $<$10 & 13 & 80 & $<$10 & 12 & 80 \\
        SG: $f_0=153$ Hz, $Q=8.9$ & $<$10 & 13 & 150 & $<$10 & 13 & 45 & $<$10 & 13 & 45 \\
        SG: $f_0=1053$ Hz, $Q=9$ & $<$10 & 11 & 63 & $<$10 & 12 & 52 & $<$10 & 11 & 52 \\
        \hline
        GA: $\tau=0.1$ ms & $<$10 & 13 & 91 & $<$10 & 13 & 60 & $<$10 & 13 & 68 \\
        GA: $\tau=2.5$ ms & $<$10 & $<$10 & 29 & $<$10 & $<$10 & 20 & $<$10 & $<$10 & 29 \\
        GA: $\tau=4.0$ ms & $<$10 & 11 & 56 & $<$10 & 11 & 56 & $<$10 & 11 & 56 \\
        \hline
        WNB: $f_0=100$ Hz, $\Delta=100$ Hz, $\tau=100$ ms & 15 & N/A & N/A & 15 & N/A & N/A & 15 & N/A & N/A \\
        WNB: $f_0=250$ Hz, $\Delta=100$ Hz, $\tau=100$ ms & 16 & N/A & N/A & 16 & N/A & N/A & 16 & N/A & N/A \\
        WNB: $f_0=1000$ Hz, $\Delta=10$ Hz, $\tau=100$ ms & 11 & 25 & N/A & 11 & 25 & N/A & 11 & 26 & N/A \\
        WNB: $f_0=1000$ Hz, $\Delta=1000$ Hz, $\tau=10$ ms & 13 & 27 & N/A & 13 & 28 & N/A & 13 & 27 & N/A\\
    \end{tabular}
\end{table*}

Finally, we explore how the signal population with which we train our signal likelihood affects our detection efficiency.  
We created three separate training populations: one consisting of only sine-Gaussians, one consisting of only white-noise bursts, and one consisting of both sine-Gaussians and white-noise bursts.  
The population of sine-Gaussians were distributed consistently with a uniform-in-volume distribution, uniformly in central frequency between 40 Hz and 1500 Hz, and uniformly in Q between 3 and 30.  
The population of white noise bursts were distributed consistently with a uniform-in-volume distribution, uniformly in starting frequency between 40 Hz and 1500 Hz, uniformly in bandwidth between 10 Hz and 1500 Hz, and uniformly in duration between 5 ms and 100 ms.  
The goal of these populations was to create an inclusive set of events to train on that intentionally had some mismatch with our LIB priors.

The detection efficiency results for all of these training scenarios are quite similar at a significance FAR of 1 per year. 
To be sure, there is some variation in $\text{SNR}_{90\%}$, but the SNRs at which this variation occurs are extremely large and probably of non-astrophysical values.  

This similarity is not surprising since the Gaussian KDE models the likelihoods well in regions of parameter space where the sample density is high, i.e., for the bulk of the distribution.  
The bulk of the distribution is able to establish the general properties of signal events as opposed to those of noise events.
Signal events tend to be louder than noise events (i.e., they have a larger BSN), and signal events tend to be more correlated than noise events (i.e., they have larger BCI).  
As seen in \S~\ref{SubSubSec.EffMorph}, the differences in oLIB's behavior for different morphologies only becomes pronounced at extremely-loud SNRs.  
These extremely-loud-SNR events are sufficiently rare for uniform-in-volume populations that their contribution to the training is negligible when compared to that of the bulk of events.  
Thus, because oLIB behaves similarly across morphologies for the bulk of events that dominate training, the likelihood models are effectively invariant to the exact morphologies used in the training.

\section{Summary}\label{Sec.Conclusions}

In this paper, we introduced the justification and methodology for a new end-to-end search algorithm targetting GW bursts called oLIB.
This algorithm takes in calibrated strain data and compresses it into a set of search statistics that can be used to make a detection statement independently of other algorithms.  
Specifically, the compression involves several steps.  
First, Omicron is used to flag stretches of excess power in each detector, which we refer to as events.  
For computational reasons, these events are down-selected by imposing constraints such as event clustering, vetoing based on data quality, and requiring a time-of-flight time coincidence across the network of detectors.  
Once this incoherent down-selection is complete, these coincident events are compressed into a set of Bayes factors with LIB, an MCMC algorithm used to calculate Bayesian evidences.  
Because Bayes factors are sufficient statistics for binary hypothesis testing, we expect the information loss concerning the signal-versus-noise hypothesis to be minimal as long as the set of oLIB's hypotheses model the actual data sufficiently well.  
We further compress this vector of Bayes factors into a scalar likelihood ratio, which preserves all of the information regarding the signal-versus-noise hypothesis contained within the set of Bayes factors.  
Finally, we use a likelihood-ratio test to assign a detection significance to each event.
This LRT allows us to optimally extract this signal-versus-noise information that we have been preserving in our compression and make a detection statement.

In order to demonstrate the validity of the algorithm's implementation, we ran oLIB over a stretch of real interferometer data taken from the initial LIGO S6 science run.  
We also injected simulated GW signals into this data in order to study the algorithm's behavior when analyzing detection candidates of varying morphology and strength.

We showed that the algorithm is capable of detecting events across a range of morphologies at astrophysically relevant SNRs.
These detection statements can be made in low-latency, on the order of tens of minutes.
We showed that, from a detection efficiency standpoint, the most powerful search involves an LRT that considered a combination of Bayes factors as search statistics.
Thus, this is the first GW burst search to optimally extract detection information from a set of multiple Bayes factors.
Finally, we confirmed that the detection efficiency of the LRT is quite robust against the exact choice of source population used when modeling the likelihoods.

The development of the oLIB unmodeled search algorithm is promising on several fronts.  
First, it provides a new end-to-end method for detecting GW bursts independently of other algorithms.  
At worst, oLIB provides overlap with existing methods that would be useful for consistency checks and validation, and, at best, oLIB provides increased sensitivity to areas of the burst parameter space.
Quantifying this overlap with existing algorithms via comparison studies has been \cite{BCP,O1AllSky} and will be an integral part of current and future joint searches for GW bursts.
Finally, since the most efficient configuration of the oLIB algorithm involves combining several search statistics through an LRT to make a detection significance statement, we have successfully demonstrated a procedure that could be used to optimally combine the search statistics across several different search algorithms into a joint detection significance statement.

\section{Acknowledgments}

The authors acknowledge the support of the National Science Foundation, the LIGO Laboratory, and the Centre National de la Recherche Scientifique (CNRS).
LIGO was constructed by the California Institute of Technology and Massachusetts Institute of Technology with funding from the National Science Foundation and operates under cooperative agreement PHY-0757058.
The authors would like to acknowledge the LIGO Data Grid clusters, without which the simulations could not have been performed. 
Specifically, we thank the Albert Einstein Institute in Hannover, supported by the Max-Planck-Gesellschaft, for use of the Atlas high-performance computing cluster.  
We would like to thank Peter Couvares and Josh Willis for help with optimizing our code.
We would also like to thank Lindy Blackburn, Kipp Cannon, Robert Eisenstein, Phillip Graff, Satya Mohapatra, Matt Pitkin, Laura Sampson, Ruslan Vaulin, Alberto Vecchio, John Veitch, and the LIGO-Virgo Burst search working group for useful comments and discussion.
This is LIGO document number P1500220.


\bibliography{refs}


\appendix

\section{Information-Theoretic Justifications}\label{App.LRTOptimal}

\subsection{Proof that Likelihood Ratios are Sufficient Statistics for Hypothesis Testing}\label{App.LRSSProof}

As mentioned in Section~\ref{SubSec.InformationJustification}, the Data Processing Inequality (see Eq.~\ref{Eq.DPI}) states the compression of a datastream $\vec{x}$ into a statistic $t(\vec{x})$ must lose information regarding the true hypothosis \Hypo unless $t(\vec{x})$ is a sufficient statistic.  
A statistic is sufficient if and only if it satisfies Eq.~\ref{Eq.SS_def}.  
It can also be shown that statistics are sufficient if and only if the likelihood $P_{x|H}(\vec{x}|H)$ can be factored into a form that satisfies the Neyman-Fisher factorization \cite{casella2002statistical}
  \begin{equation}
  P_{x|H}(\vec{x}|H) = a(t(\vec{x}),H)b(\vec{x})
  \end{equation}
where $a$ can be a function only of $t(\vec{x})$ and \Hypo and $b$ can only be a function of $\vec{x}$.

Using the Neyman-Fisher factorization, we can show that the likelihood ratio $\Lambda(\vec{x})$ is a sufficient statistic with respect to $P_{x,H}(\vec{x},H)$ where $\vec{x}$ is a random vector of analysis statistics and $H \in \{H_0,H_1\}$ is a random hypothesis variable for binary hypothesis testing.  In order to prove this statement, we consider the form of $P_{x|H}(\vec{x}|H)$ under both hypotheses:
    \begin{equation}
    P_{x|H}(\vec{x}|H=H_1) = \Lambda(\vec{x}) \cdot P_{x|H}(\vec{x}|H=H_0)
    \end{equation}
    \begin{equation}
    P_{x|H}(\vec{x}|H=H_0) = 1 \cdot P_{x|H}(\vec{x}|H=H_0)
    \end{equation}
where $\Lambda(\vec{x}) \equiv \frac{P_{x|H}(\vec{x}|H=H_1)}{P_{x|H}(\vec{x}|H=H_0)}$ is the likelihood ratio.  
Thus, by defining $\Lambda_{i,j}(\vec{x}) \equiv \frac{P_{x|H}(\vec{x}|H=H_i)}{P_{x|H}(\vec{x}|H=H_j)}$, $a(\Lambda(\vec{x}), H_i) = \Lambda_{i,0}(\vec{x})$, and $b(\vec{x}) = P_{x|H}(\vec{x}|H=H_0)$, we can complete the Neyman-Fisher factorization, proving that $\Lambda(\vec{x})$ is a sufficient statistic for binary hypothesis testing.

It is straightforward to generalize this proof from binary hypothesis testing to N-dimensional hypothesis testing where $N \geq 2$.  
Here, let our statistic be $\vec{\Lambda}(\vec{x})$, the set of all possible likelihood ratios among the N hypotheses $\vec{H} = \{H_0, H_1, ..., H_N\}$.  
More formally, $\vec{\Lambda}(\vec{x}) = \{...,\Lambda_{i,j}(\vec{x}),...\}$ for all $i,j \in \{0,1,...,N\}$.  
We can then write down the form for $P_{x|H}(\vec{x}|H)$ for any arbitrary hypothesis $H_i$:
    \begin{equation}
    P_{x|H}(\vec{x}|H=H_i) = \Lambda_{i,0}(\vec{x}) \cdot P_{x|H}(\vec{x}|H=H_0) .
    \end{equation}
Thus, by defining $a(\vec{\Lambda}(\vec{x}), H) = \Lambda_{i,0}(\vec{x})$ and $b(\vec{x}) = P_{x|H}(\vec{x}|H=H_0)$, we see that we can complete the Neyman-Fisher factorization, proving that the set of all likelihood ratios $\vec{\Lambda}(\vec{x})$ is a sufficient statistic for N-dimensional hypothesis testing.  Actually, closer inspection shows that we do not even need $\vec{\Lambda}(\vec{x})$ to be the full set of likelihood ratios between all hypotheses, but rather only the set of likelihood ratios between all individual hypotheses and a particular hypothesis $H_0$ (since the likelihood ratio of any two arbitrary hypotheses $H_i$ and $H_j$ can be computed through the ratio $\Lambda_{i,j}(\vec{x}) = \frac{\Lambda_{i,0}(\vec{x})}{\Lambda_{j,0}(\vec{x})}$.

This property implies that the set of likelihood ratios spanning all possible hypotheses $\vec{\Lambda}(\vec{x})$ is optimal in a data processing sense.  
As shown by the Data Processing Inequality, a statistic that is sufficient with respect to a random data vector $\vec{x}$ and a random variable \Hypo preserves all of the mutual information shared between those two variables.  
Thus, in N-dimensional hypothesis testing, no information about the actual hypothesis \Hypo is lost when compressing the statistic vector $\vec{x}$ into the set of likelihood ratios $\vec{\Lambda}(\vec{x})$.  
While other combinations of $\vec{x}$ may also be sufficient statistics (sufficient statistics are not inherently unique), they cannot contain more information about \Hypo than $\vec{\Lambda}(\vec{x})$ does, and thus $\vec{\Lambda}(\vec{x})$ is an optimal statistic in N-dimensional hypothesis testing.

On a final note, let us consider the case where all of the specific hypotheses \Hypo in the N-dimensional $\vec{H}$ can be categorized as a element of a greater positive-versus-null binary hypothesis $H_{\text{bin}}$, so that $H \in H_{\text{bin}} \in \{H_{\text{null}},H_{\text{pos}}\}$ (e.g., if $\vec{H}$ contains different signal and noise models, but all are sub-models of the greater signal or noise hypotheses $H_{\text{signal}}$ and $H_{\text{noise}}$). 
In this scenario, assuming that the true hypothesis is an element of $\vec{H}$, the original data vector $\vec{x}$ can be compressed into the set of likelihood ratios spanning all possible hypotheses $\vec{\Lambda}(\vec{x})$, and this compression is lossless with respect to the information concerning the specific hypothesis \Hypo.  
We can show that this compression is also lossless with respect to the information concerning the binary hypothesis $H_{\text{bin}}$ by considering the likelihoods for both the null and positive binary hypotheses:
    \begin{equation}
    \begin{split}
    P_{x|H_{\text{bin}}} & (\vec{x}|H_{\text{null}}) \\
			 &= \sum_{i} P_{x|H,H_{\text{bin}}}(\vec{x}|H_i,H_{\text{null}})P_{H|H_{\text{bin}}}(H_{i}|H_{\text{null}}) \\
                         &= P_{x|H,H_{\text{bin}}}(\vec{x}|H_0,H_{\text{null}}) \\
                         &\quad \times \sum_{i} \Lambda_{i,0}(\vec{x}) P_{H|H_{\text{bin}}}(H_{i}|H_{\text{null}})
    \end{split}
    \end{equation}
    \begin{equation}
    \begin{split}
    P_{x|H_{\text{bin}}} & (\vec{x}|H_{\text{pos}}) \\
			 &= \sum_{i} P_{x|H,H_{\text{bin}}}(\vec{x}|H_i,H_{\text{pos}})P_{H|H_{\text{bin}}}(H_{i}|H_{\text{pos}}) \\
                         &= P_{x|H,H_{\text{bin}}}(\vec{x}|H_0,H_{\text{null}}) \\
                         &\quad \times \sum_{i} \Lambda_{i,0}(\vec{x}) P_{H|H_{\text{bin}}}(H_{i}|H_{\text{pos}})
    \end{split}
    \end{equation}
By defining $a(\vec{\Lambda}(\vec{x}), H_{\text{bin}})$ to be the correct positive-versus-null hypothesis summation term and $b(\vec{x}) = P_{x|H,H_{\text{bin}}}(\vec{x}|H_0,H_{\text{null}})$, we complete the Neyman-Fisher factorization and show that $\vec{\Lambda}(\vec{x})$ is a sufficient statistic with respect to the information contained within $\vec{x}$ about $H_{\text{bin}}$ in addition to being a sufficient statistic with respect to \Hypo\footnote{We note that the Neyman-Fisher factorization is still satisfied if we define the statistic $t(\vec{x})$ to be the correct positive-versus-null summation term.  
Thus, the set of expectation values over all likelihood ratios with respect to \textit{both} the signal and noise hypothesis likelihoods $P_{H_i|H_\text{bin}}(H_i|H_\text{bin})$ is a sufficient statistic with respect to the binary hypotheis $H_\text{bin}$.}.  
Thus, if we further losslessly compress $\vec{\Lambda}(\vec{x})$ into a single, scalar search statistic $\Lambda_{\text{bin}}(\vec{\Lambda}(\vec{x}))$, we have not lost any information about $H_{\text{bin}}$.  
\textit{This is an important result}.  
It shows that we can achieve lossless data compression into a scalar in two ways:  1.) directly into $\Lambda_{\text{bin}}$ from the data vector $\vec{x}$, or 2.) first into a set of likelihood ratios spanning a set of embedded sub-hypotheses and then into $\Lambda_{\text{bin}}$.

\subsection{Adding Search Statistics can only Increases Information}\label{App.InfoMultVars}
Any vector of search statistics $\vec{x}$ contains a non-negative amount of information $I(H;\vec{x})$ about a hypothesis \Hypo.  
Eq.~\ref{Eq.IdefH} demonstrates that this mutual information can be interpreted as the reduction in entropic uncertainty of \Hypo achieved by having knowledge of the search statistics $\vec{x}$.
By explicitly defining the entropy $\mathcal{H}(\cdot)$ and conditional entropy $\mathcal{H}(\cdot|\cdot)$ to be
    \begin{equation}\label{Eq.Hdef}
    \mathcal{H}(a) = -\sum_{a'} P_a(a') \log P_a(a')
    \end{equation}
    \begin{equation}\label{Eq.HCondef}
    \mathcal{H}(a|b) = -\sum_{a',b'} P_{a,b}(a',b') \log P_{a|b}(a'|b')
    \end{equation}
we can also define the mutual information $I(\cdot;\cdot)$ and conditional mutual information $I(\cdot;\cdot|\cdot)$ to be
    \begin{equation}\label{Eq.IdefP}
    I(a;b) = \sum_{a',b'} P_{a,b}(a',b') \log \frac{P_{a,b}(a',b')}{P_a(a')P_b(b')} \ \ .
    \end{equation}
    \begin{equation}\label{Eq.ICondefP}
    I(a;b|c) = \sum_{a',b',c'} P_{a,b,c}(a',b',c') \log \frac{P_{a,b|c}(a',b'|c')}{P_{a|c}(a'|c')P_{b|c}(b'|c')} \ \ .
    \end{equation}
    
It is interesting to study what happens to the mutual information when we change the dimension of $\vec{x}$, i.e., what happens when we add or remove a given search statistic from our vector.  
We can consider the mutual information with an added search statistic $x_+$ by explicitly writing out $I(H;\vec{x},x_+)$ and factoring the probabilities:
  \begin{equation}
  I(H;\vec{x},x_+) = \sum_{H,x,x_+} P_{H,x}(H,\vec{x},x_+) \log \frac{P_{H,x}(H,\vec{x},x_+)}{P_{x}(\vec{x},x_+) P_{H}(H)}
  \end{equation}
  \begin{equation}
  \begin{split}
  I(H;\vec{x},x_+) &= \sum_{H,x,x_+} P_{H,x}(H,\vec{x},x_+) \\
		   &\quad \times \log \frac{P_{H,x}(H,\vec{x}) \cdot P_{x|H,x}(x_+|H,\vec{x}) P_{H|x}(H|\vec{x})}{P_{x}(\vec{x}) P_{H}(H) \cdot P_{x|x}(x_+|\vec{x}) P_{H|x}(H|\vec{x})}
  \end{split}
  \end{equation}
  \begin{equation}
  I(H;\vec{x},x_+) = I(H;\vec{x}) + I(H;x_+|\vec{x})  .
  \end{equation}
 
We can write any conditional mutual information $I(a;b|c)$ as
  \begin{equation}
  \begin{split}
  I(a;b|c) = &\sum_c P_c(c') \sum_{a,b} [ P_{a,b|c}(a',b'|c') \log P_{a,b|c}(a',b'|c') \\
	     &- P_{a,b|c}(a',b'|c') \log P_{a|c}(a'|c') P_{b|c}(b'|c') ] \ \ .
  \end{split}
  \end{equation}
A straightforward application of the Gibbs inequality and the non-negativity of probabilities makes it possible to show $I(a;b|c) \geq 0$.  
Thus we have
  \begin{equation}
  I(H;\vec{x},x_+) \geq I(H;\vec{x})  
  \end{equation}
which proves that adding a search statistic can only add to the mutual information, and thus it can only decrease the entropic uncertainty $\mathcal{H}(H|\vec{x})$ of that hypothesis.  In other words, adding a search statistic can only make $P_{H|x}(H|\vec{x})$ a more sharply-peaked distribution.


\section{Gaussian Kernel Density Estimation}\label{App.KDE}

In order for us to use an LRT for our signal-versus-noise binary hypothesis test, we need models of the signal and noise likelihoods.  Without a given functional form for these likelihood distributions, we must find a way of approximating them in some optimal sense.  The Kullback-Leibler divergence between two distributions P and Q, defined as
  \begin{equation}\label{Eq.KL}
  D(P||Q) = \sum_{i} P(i) \log \frac{P(i)}{Q(i)}  ,
  \end{equation}
provides a measure of the distance between two distributions.  It represents the reduction in entropy when using the true distribution $P$ instead of the wrong distribution $Q$, or in an information-theoretic sense, it measures the loss of information when using the wrong distribution $Q$ instead of the true distribution $P$.  Thus, if we wish to model the true distribution $f(\vec{x})$ of our search statistics $\vec{x}$ with a model distribution $\hat{f}(\vec{x})$, we should minimize $D(f||\hat{f})$ in order to maximize the information that is contained within $\hat{f}(\vec{x})$ about $\vec{x}$.  By changing the sum in Eq.~\ref{Eq.KL} to an integral in order to account for continuous variables, the quantity to be minimized becomes
  \begin{equation}\label{Eq.KL_opt_1}
  D(f||\hat{f}) = \int f(\vec{x}) \log f(\vec{x}) d\vec{x} - \int f(\vec{x}) \log \hat{f}(\vec{x}) d\vec{x}  .
  \end{equation}
Since only the second term in Eq.~\ref{Eq.KL_opt_1} depends on our model choice $\hat{f}(\vec{x})$, the optimization problem becomes a maximization of
  \begin{equation}\label{Eq.KL_opt_2}
  B = \int f(\vec{x}) \log \hat{f}(\vec{x}) d\vec{x}  .
  \end{equation}

With our optimization criterion in place, we must then choose our model $\hat{f}(\vec{x})$.  One non-parametric approach to this problem is that of kernel density estimation (KDE).  KDE consists of centering an N-dimensional kernel at each of a set of N-dimensional empirical data points drawn from $f(\vec{x})$.  These kernels are then summed over, and the normalized sum is used as the distribution model $\hat{f}(\vec{x})$.  When identical Gaussian kernels are used for each data point, this model takes the form
  \begin{equation}\label{Eq.KDE}
  \hat{f}(\vec{x}) = \frac{1}{n \sqrt{(2\pi)^N |H|}} \sum_i^n e^{-\frac{1}{2}(\vec{x}-\vec{d}_i)^\intercal H^{-1} (\vec{x}-\vec{d}_i)}
  \end{equation}
where $i$ indexes one of n data points drawn from the true distribution $f(\vec{x})$ and H is a matrix representing the squared bandwidths of the kernels.  A kernel's bandwidth $h_m$ controls the width of the kernel (i.e., the extent to which it models local versus distant parts of the parameter space) in the $m^{th}$ dimension.  If we choose all of the N bandwidths (one for each dimension) to be uncorrelated, then H is a diagonal matrix with $h_m^2$ as the $m^{th}$ entry along the diagonal.

In order to evaluate Eq.~\ref{Eq.KL_opt_2}, we need to know the functional form of $f(\vec{x})$.  We can approximate this using the empirical approximation
  \begin{equation}
  \int f(\vec{x}) g(\vec{x}) d\vec{x} = E_f[g(\vec{x})] \approx \frac{1}{n} \sum_{j}^{n} g(\vec{d}_j)
  \end{equation}
where we replace the integral over $\vec{x}$ with a sum over the n data points ${\vec{d}_1,...,\vec{d}_n}$ sampled from $f(\vec{x})$.  For our purposes, $g(\vec{x}) = \log \hat{f}(\vec{x})$, giving us
  \begin{equation}
  B \approx \frac{1}{n} \sum_j^n \log \left( \frac{1}{n \sqrt{(2\pi)^N |H|}} \sum_i^n e^{-\frac{1}{2}(\vec{d}_j -\vec{d}_i)^\intercal H^{-1} (\vec{d}_j-\vec{d}_i)} \right)  .
  \end{equation}
Finally, in order to prevent ourselves from overtraining the data, we use leave-one-out cross-validation by removing the $j^{th}$ data point from the inner sum, yielding the expression
  \begin{widetext}
  \begin{equation}\label{Eq.KDE_opt_final}
  B \approx \frac{1}{n} \sum_j^n \log \left( \frac{1}{(n-1) \sqrt{(2\pi)^N |H|}} \sum_{i \neq j}^n e^{-\frac{1}{2}(\vec{d}_j -\vec{d}_i)^\intercal H^{-1} (\vec{d}_j-\vec{d}_i)} \right)  .
  \end{equation}
  \end{widetext}
The result of overtraining can be seen by considering the case where $H \rightarrow \bold{0}$.  In this limit, all of the Gaussian Kernels become Dirac delta functions centered around $d_j$.  Thus, the $i=j$ point provides an infinite contribution to B, meaning a zero-bandwidth KDE is optimal and that the optimal estimate of $f(\vec{x})$ is simply the set of empirical data points.  Removing the $i=j$ point from the sum helps prevent this overtraining, although it should be noted that the zero-bandwidth B will be infinite and therefore maximal if any of the data points are exact duplicates (which becomes more and more unlikely as the dimensionality N increases).

In practice, we find the optimal bandwidths of our KDE likelihood estimates by maximizing Eq.~\ref{Eq.KDE_opt_final} over a grid in the N-dimensional parameter space.  In cases where the zero bandwidth is infinite, we search instead for a secondary local maximum.

\end{document}